\documentclass[]{article}

\usepackage{savesym}
\usepackage{graphicx}

\usepackage{lineno}

\usepackage{xcolor}

\savesymbol{tablenum}
\usepackage{siunitx}

\restoresymbol{SIX}{tablenum}

\usepackage[version=3]{mhchem}
\usepackage{makecell}

\usepackage{geometry}
\usepackage[utf8]{inputenc}
\usepackage[T1]{fontenc}

\usepackage{amsmath}
\usepackage{amssymb}
\usepackage{mathrsfs}

\usepackage{lineno}

\usepackage{tikz}
\usetikzlibrary{plotmarks,patterns,calc,decorations.markings}

\usepackage{booktabs}

\usepackage{lipsum}

\usepackage[symbol]{footmisc}

\catcode`\@=11 
\def\parenbar{\mathpalette\p@renb@r}
\def\p@renb@r#1#2{\vbox{%
\ifx#1\scriptscriptstyle \dimen@.7em\dimen@ii.2em\else
\ifx#1\scriptstyle \dimen@.8em\dimen@ii.25em\else
\dimen@1em\dimen@ii.4em\fi\fi \offinterlineskip
\ialign{\hfill##\hfill\cr
\vbox{\hrule width\dimen@ii}\cr
\noalign{\vskip-.3ex}%
\hbox to\dimen@{$\mathchar300\hfil\mathchar301$}\cr
\noalign{\vskip-.3ex}%
$#1#2$\cr}}}
\catcode`\@=12 
\def\nuan{\parenbar{\nu}\kern-0.4ex}

\newlength{\smfigwidth}
\newlength{\figwidth}
\newlength{\captwidth}
\usepackage[pdflatex]{trimclip}

\usepackage{hyperref}
\usepackage[all]{hypcap} 
\usepackage{authblk}

\date{\today}

\title{Search for neutrino counterparts of catalogued gravitational-wave events detected by Advanced-LIGO and Virgo during run O2 with ANTARES}

\author[1,2]{A.~Albert}
\author[3]{M.~Andr\'e}
\author[4]{M.~Anghinolfi}
\author[5]{G.~Anton}
\author[6]{M.~Ardid}
\author[7]{J.-J.~Aubert}
\author[8]{J.~Aublin}
\author[8]{B.~Baret}
\author[9]{S.~Basa}
\author[10]{B.~Belhorma}
% ...

\affil[1]{\scriptsize{Universit\'e de Strasbourg, CNRS,  IPHC UMR 7178, F-67000 Strasbourg, France}}
\affil[2]{\scriptsize{Universit\'e de Haute Alsace, F-68200 Mulhouse, France}}
\affil[3]{\scriptsize{Technical University of Catalonia, Laboratory of Applied Bioacoustics, Rambla Exposici\'o, 08800 Vilanova i la Geltr\'u, Barcelona, Spain}}
\affil[4]{\scriptsize{INFN - Sezione di Genova, Via Dodecaneso 33, 16146 Genova, Italy}}
\affil[5]{\scriptsize{Friedrich-Alexander-Universit\"at Erlangen-N\"urnberg, Erlangen Centre for Astroparticle Physics, Erwin-Rommel-Str. 1, 91058 Erlangen, Germany}}
\affil[6]{\scriptsize{Institut d'Investigaci\'o per a la Gesti\'o Integrada de les Zones Costaneres (IGIC) - Universitat Polit\`ecnica de Val\`encia. C/  Paranimf 1, 46730 Gandia, Spain}}
\affil[7]{\scriptsize{Aix Marseille Univ, CNRS/IN2P3, CPPM, Marseille, France}}
\affil[8]{\scriptsize{Universit\'e de Paris, CNRS, Astroparticule et Cosmologie, F-75013 Paris, France}}
\affil[9]{\scriptsize{Aix Marseille Univ, CNRS, CNES, LAM, Marseille, France }}
\affil[10]{\scriptsize{National Center for Energy Sciences and Nuclear Techniques, B.P.1382, R. P.10001 Rabat, Morocco}}
% ...

\begin{document} 


\maketitle 

\begin{abstract}
An offline search for a neutrino counterpart to gravitational-wave (GW) events detected during the second observation run (O2) of Advanced-LIGO and Advanced-Virgo performed with ANTARES data is presented. In addition to the search for long tracks induced by $\nu_\mu$ ($\overline{\nu}_{\mu}$) charged current interactions, a search for showering events induced by interactions of neutrinos of any flavour is conducted. The severe spatial and time coincidence provided by the gravitational-wave alert allows regions above the detector horizon to be probed, extending the ANTARES sensitivity over the entire sky. The results of this all-neutrino-flavour and all-sky time dependent analysis are presented. The search for prompt neutrino emission within $\pm$500~s around the time of six GW events yields no neutrino counterparts. Upper limits  on the neutrino spectral fluence and constraints isotropic energy radiated via high-energy neutrinos (from a few TeV to a few tens of PeV) are set for each GW event analysed.
\end{abstract}

\section{\label{s:intro} Introduction}

Three years after the first Gravitational-Wave (GW) detection in 2015~\cite{ligo150914}, a catalogue of GW sources observed by the LIGO Scientific and Virgo collaborations has been released (Dec. 2018), spanning the entirety of the scientific runs O1 (from September 2015 until January 2016) and O2 (from November 2016 until August 2017)~\cite{GWTC-1}. Among the eleven catalogued events, four were announced for the first time, while the others were already published~\cite{ligo150914,ligo151226,ligo170104,ligo170608,ligo170814,BNS}. All of them are binary black-hole mergers (BBHs), with the only exception being the binary neutron star merger (BNS), GW170817~\cite{BNS}.

The ANTARES Collaboration actively follows all the GW alerts and performs a search for neutrino emission from GW sources. Results from the real-time and refined offline neutrino searches have been published for five GW signals~\cite{GW150914,GW151226,GW170104,GW170817}. The distinction between these two searches is that online searches only rely on upward-going muon neutrinos while offline searches have been extended to the full sky~\cite{GW170104} and to neutrinos of all flavours~\cite{GW170817}.

In this paper, results from a dedicated full-sky search for all-flavour neutrinos associated with the remaining six GW events (listed in Table~\ref{table:areas}) are presented. All the signals analysed correspond to the coalescence of binary black-hole systems. As estimated by GW detectors, the probability of these events not being astrophysical is very low, with a false alarm rate (FAR) lower than once per 70,000 yr (upper limit from gstLAL search algorithm~\cite{gstLAL}), except for GW170729 whose FAR is 0.18~yr$^{-1}$ for the same search pipeline. Their estimated distances range from 300 Mpc to about 3000 Mpc and present chirp masses from 8 M$_{\odot}$ to 36 M$_{\odot}$. With two interferometers taking data during most of O2 (November 2016 -- July 2017) and three in August 2017, the 90\% confidence level (CL) localization regions estimated by triangulation, hereafter also referred to as error box, range from 40 to 900 deg$^{2}$~\cite{GWTC-1}. 

\begin{table}[!h]
\begin{center}
  \caption{Properties of the six GW events analysed in this work. The size of the 90\% CL error box viewed as downgoing and upgoing events for ANTARES at the time of the alert is provided in the first two columns. The GCN circular numbers associated with the Advanced LIGO-Virgo and ANTARES real-time results are also provided. The last two columns  report the estimated luminosity distance and the chirp mass of each event. }
  \vspace{-0.1cm} 
  \label{table:areas}
\begin{tabular}{|c|c|c|c|c|c|}\hline   
{\bf Event} & {\bf A$_{\rm up}^{90\%}$ ($^{\circ}$)$^2$} & {\bf A$_{\rm down}^{90\%}$ ($^{\circ}$)$^2$} & {\small \#GCN GW/ANTARES} & {\bf $D_L$ [Mpc]} & {\bf $M_{\rm chirp}$ [M$_{\odot}$]} \\ \hline
GW170608 & 226 & 170 & 21221/21223 & 320$^{+120}_{-110}$ & 7.9$^{+0.2}_{-0.2}$ \\ \hline
GW170729 & 553 & 475 & No GCN sent & 2750$^{+1350}_{-1320}$ & 35.7$^{+6.5}_{-4.7}$ \\ \hline
GW170809 & 245 & 95 & 21431/21433 & 990$^{+320}_{-380}$ & 25.0$^{+2.1}_{-1.6}$ \\ \hline
GW170814 & 87 & 0 & 21474/21479 & 580$^{+160}_{-210}$ & 24.2$^{+1.4}_{-1.1}$ \\ \hline
GW170818 & 0 & 39 & No GCN sent & 1020$^{+430}_{-360}$ & 26.7$^{+2.1}_{-1.7}$ \\ \hline
GW170823 & 878 & 771 & 21656/21659 & 1850$^{+840}_{-840}$ & 29.3$^{+4.2}_{-3.2}$ \\ \hline
  \end{tabular}
 \end{center}
\end{table}


Current modeling of BBH mergers allows for an electromagnetic (EM) or neutrino counterpart in certain circumstances. If the pre-merger system is surrounded by an accretion disk, a relativistic jet could be formed upon merger, leading to proton acceleration. Particle acceleration in these jets will lead to EM radiation and potential high-energy neutrino (HEN) production~\cite{BBH,BBH2}. If the binary system is located close to an AGN, the surrounding dense material may provide the hadronic environment needed for neutrino production~\cite{bartos}. This motivates the search for a multi-messenger counterpart to BBH events.

Near-real-time alerts were issued for nine of the eleven GWs after the identification of the events. Two events did not pass the thresholds of the real-time analysis by LIGO-Virgo and therefore did not trigger an alert. For those events for which an alert was received, the ANTARES Collaboration performed rapid follow-up observations, looking for neutrino counterparts below the horizon of the ANTARES detector, i.e.\ upgoing muon neutrino candidates having travelled through the Earth.

The motivation for a real-time search of a neutrino counterpart to GW events is to promptly reduce the GW error box. The good angular resolution of ANTARES (0.4$^{\circ}$ for muon neutrinos with E$_\nu>$10 TeV \cite{PS2012}) would provide, in the case of an associated muon neutrino detection, a fast and precise localisation of the source, and allow for an efficient EM follow-up of the event. No neutrinos were observed in coincidence within a time window of $\pm$1~h around the GW event time for any of the events that triggered an alert. The results for the online search of the events that triggered a follow-up were distributed to the follow-up community via GCN circulars (numbers reported in Table~\ref{table:areas}).



Selecting upgoing events allows for an efficient rejection of the atmospheric muon background in the detector, since only neutrinos can travel through the Earth. The search for space/time coincidences between neutrinos and GW detections provides a significant background reduction that also allows to search for events above the ANTARES detector horizon, seen as downgoing in the detector frame, where the atmospheric muon background is largely dominant.

Upgoing neutrino-induced muons (or "tracks") are the main detection channel for high-energy neutrinos with the ANTARES telescope. The latest point-source and diffuse neutrino searches \cite{Giulia,Luigi} by the ANTARES Collaboration illustrate the improvement achieved with an all-flavour analysis ($\sim$30\% gain in sensitivity), i.e.\ by including also the so called "shower" event topology. This topology covers hadronic cascades induced by all-flavour neutral current interactions and electromagnetic cascades from electron and tau neutrino charged current interactions.  

In this work, a full-sky and all-flavour search for prompt neutrino emission associated with the six GW O2 events reported in Table~\ref{table:areas} is conducted. The two search regions and two neutrino topologies account for four event classes analysed separately: upgoing tracks, upgoing showers, downgoing tracks and downgoing showers.

Moreover, the updated skymaps produced with the \textit{LALInference} \cite{LALinference} reconstruction algorithm are used to evaluate the 90\% CL GW localisation contours. This is combined with the most recent ANTARES dataset, after incorporating dedicated calibrations \cite{calib1,calib2,calib3}, leading to an improved reconstruction. The search time window of $\pm$500~s around the GW event has been chosen according to ref.~\cite{baret}. The analysis is optimised for a 3$\sigma$ significance in case an event is observed in space and time coincidence with the GW signal.

This document is organised as follows. The characteristics of the GW events analysed are given in Sect.~\ref{s:events}. The neutrino event selection is detailed in Sect.~\ref{s:opt} for each of the four event classes considered. In Sect.~\ref{s:results}, the main results and astrophysical constraints are presented and discussed. Finally, the conclusions and perspectives are highlighted in Sect.~\ref{s:conc}.  

\section{\label{s:events} The GW catalogued events}

The GW data analysis of the events in the catalogue was carried out in parallel with three different pipelines by the LIGO/Virgo Collaboration. Two pipelines use cross-correlation with signal templates specifically for compact binary mergers (GstLAL~\cite{gstLAL} and pyCBC~\cite{pyCBC}). The third is independent on the signal pattern and uses a time-frequency analysis to search for a transient and unmodelled GW signal, coherent Wave Burst (cWB~\cite{cWB}). Some information about these events is given below: \\
- GW170608 happened during a special period in which the LIGO-Hanford detector was going through a process for angular noise stabilization while the LIGO-Livingston detector was operating in a nominal configuration.\\ 
- GW170729 is interesting because it was found with the highest significance by the weakly-modelled searches\\ (cWB~\cite{cWB}) and one of the merging black holes presents a reconstructed mass which is beyond the predicted limits from stellar evolution. This possibly points towards a candidate of a different astrophysical origin. It was only identified in the offline search and thus, no alert was sent to EM observatories. It is also the GW event with the most distant origin, the most massive black hole remnant and the only one for which a null post-merger spin can be ruled out.\\  
- GW170809 was found online and thus triggered an alert that was sent to EM partners but that did not pass the offline selections.\\
- GW170814 was identified as a double-coincident event between Livingston and Hanford detectors by GstLAL when re-analyzing O2 data to incorporate an updated calibration on the Virgo data, and the noise subtraction in the LIGO data. It is also well localised on the sky thanks to the non-observation by Virgo.\\
- GW170818 occurred just one day after the binary neutron star merger (GW170817). It did not pass all the online triggers and thus no alert was sent for an EM counterpart search. It was just found by GstLAL in the offline analysis. It is the binary black hole merger with the best reconstructed location up to now.\\  
- GW170823 was triggered by the three GW pipelines online and offline, but with low significance before the updated analysis.

\section{\label{s:opt} Analysis method}

A blind search for prompt neutrino emission correlated with the GW signal is performed. For all the four event classes analysed, the background expectation inside the error box of each GW event is inferred directly from data outside the search window. Without selection cuts, the reconstructed data is largely dominated by the atmospheric muon background, while neutrino induced events are better reconstructed than atmospheric muons and will remain after the cuts. Since the region in the parameter space where the selection cut is applied is sparsely populated in data (about two events per day with the selection criteria used to search for persistent point sources~\cite{PS2012}), this contribution is scrutinised by using a dedicated run-by-run Monte Carlo (MC) simulation reproducing the data-taking conditions of the ANTARES detector at the time of each GW alert~\cite{mcrbr}. 

For the estimation of the number of expected background events within 1000~s for each considered run, the background rate is assumed to be uniform in time over the run duration (of about 12h)~\cite{GW170104}. The estimated background level inside the error box and within the 1000~s window is further reduced by applying analysis cuts on the quality parameters of the event reconstruction. The values of the cuts are chosen so that the detection of one event in coincidence with the GW would correspond to a 3$\sigma$ discovery. For each individual GW candidate, the optimised value of the cut (or set of cuts) is defined as the value(s) for which the expected number of selected background events in time and space coincidence with the GW is such that the Poisson probability of observing at least one coincident background event becomes smaller than $p_{3\sigma} = 2.7 \times 10^{-3}$.


\subsection{Track event selection}

The track event selection procedure is described in ref.~\cite{GW170104}. Different optimisations are used for events coming from below and above the horizon. 

For upgoing events, the optimisation is model independent and it is done only on the reconstruction quality parameter ($\Lambda$), computed as the ratio between the reconstruction likelihood and the number of degrees of freedom \cite{PS2012}.

For downgoing events, the energy estimate of the events is also used to further reduce the overwhelming atmospheric muon background. The number of hits used in the event reconstruction (N$_{\rm hits}$) is used as a proxy of the energy estimate. For each GW event, the set of cuts on $\Lambda$ and N$_{\rm hitts}$ that maximises the number of detected signal events in the arrival direction likelihood map assuming a $\frac{dN}{dE} \propto E^{-2}$ neutrino spectrum, is chosen among those that fulfill the $p_{3\sigma}$ condition.


\subsection{Shower event selection}

All events triggered in the ANTARES detector that do not pass the track event selection described above and that are reconstructed by the ANTARES shower reconstruction algorithm~\cite{TANTRA} enter in the shower sample. As a consequence, the track and shower samples are disjoint. Additionally, only events that are contained in the detector are considered as shower-like candidates. The containment is defined in ref.~\cite{PS2012}.

After the pre-selection (based, as in ref.~\cite{GW170817}, on a reconstruction parameter called M-estimator), two additional parameters are used to optimise the shower sample. The optimisation procedure is the same for upgoing and downgoing events. The result of a Random Decision Forest (RDF) classifier is used to distinguish track-like events from shower-like events. In addition, an extended likelihood ratio, $L_{\mu}$, is also used to discriminate between cosmic showers and atmospheric muons based on the photomultiplier hits information~\cite{lmu}.

For each cut in the RDF output, the optimised $L_{\mu}$ cut is obtained, using the same $p_{3 \sigma}$ criterion as for tracks. The set of cuts on RDF and $L_{\mu}$ that maximizes the surviving signal assuming an $E^{-2}$ neutrino spectrum while fulfilling the $p_{3 \sigma}$ condition is chosen as the final selection.

\section{\label{s:results} Search results and astrophysical constraints}

No neutrinos coincident in space and time with any of the GW signals analysed have been found after unblinding the dataset. The non-detection of a transient neutrino signal from the catalogued GW sources is used to set constraints on the neutrino emission. 

Given that the detector sensitivity depends on the source position, and that there is no precise location of the GW signals, the neutrino emission from a point-like source will be considered, with the source located at different pixels inside the error box region. The pixel size in which the detector sensitivity is computed is chosen to be large enough to avoid MC statistical fluctuations (18$^{\circ}$ in declination and 36$^{\circ}$ in right ascension). Upper limits (UL) as a function of the position on the sky are presented in the form of skymaps (see Fig.~\ref{fig:limits}). Constraints are set both on the neutrino spectral fluence and on the total isotropic energy emitted through high-energy neutrinos in the 5-95\% energy range sensitivity of the search, with results summarized in Tables~\ref{table:limits} and~\ref{tab:eiso}.

\subsection{Constraints on the neutrino spectral fluence}

Upper limits at 90\% CL on the neutrino spectral fluence from a point-like source located in a given position on the sky are calculated using the null result and the detector acceptance.

The ANTARES acceptance, i.e. the number of selected signal events per given unit flux, and effective area are evaluated by means of a dedicated MC simulation performed on a run-by-run basis. This includes the detector configuration and variable data-taking conditions for each ANTARES observing run at the time of the GW events under study. 

In the case of no signal event, a 90\% CL fluence upper limit can be defined. Using Poisson statistics, this upper limit corresponds to the time integrated flux that would produce on average $N^{90\%}$=2.3 detected neutrino candidates in the pixel containing the source. The 90\% UL on the number of events, $N^{90\%}$, is defined as:
\begin{equation}
    N^{90\%} = \int \frac{dN}{dE_{\nu}}(E_{\nu},\delta) A_{\rm eff}(E_{\nu},\delta) dE_{\nu}
\end{equation}
where $A_{\rm eff}(E_{\nu},\delta)$ is the ANTARES effective area at the alert time, which takes into account the absorption of neutrinos by the Earth and the detector visibility. This effective area depends on the event selection cuts as well as on the neutrino energy and the position of the source. For a neutrino power-law spectrum ($\frac{dN}{dE_{\nu}} \propto E^{-\gamma}$), the spectral fluence at the detector can be defined as:
\begin{equation}
    E_{\nu}^{2} \frac{dN}{dE_{\nu}} = \phi_{0} \left( \frac{E_{\nu}}{\rm 1 GeV} \right)^{-\gamma + 2} {\rm (GeV \cdot cm^{-2})}.
\end{equation}

\begin{figure}
    \vspace{-3.5cm}

    \hspace{-1.2cm}\includegraphics[scale=0.505]{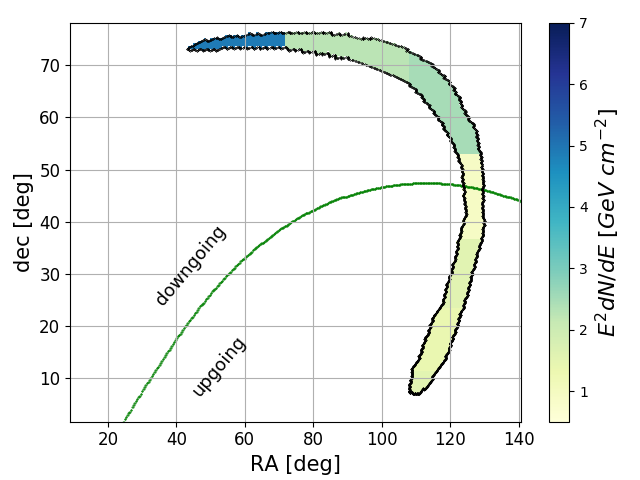}\hspace{0.15cm}
    \includegraphics[scale=0.51]{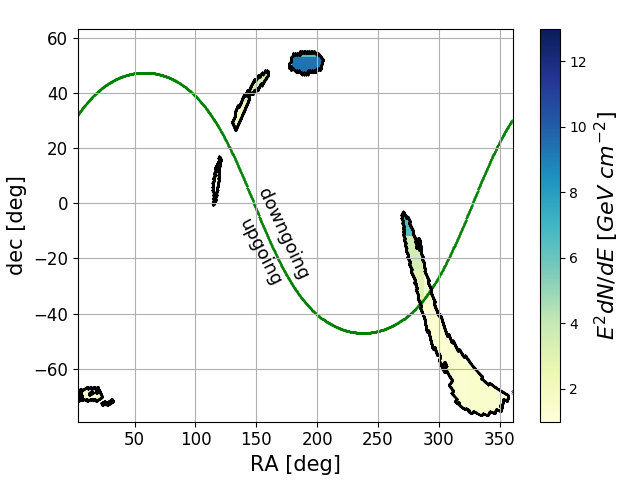}
    
    \hspace{-1.2cm}\includegraphics[scale=0.52]{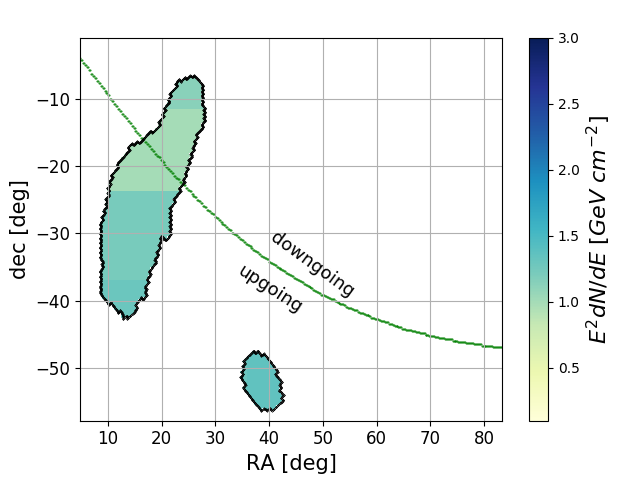}
    \includegraphics[scale=0.52]{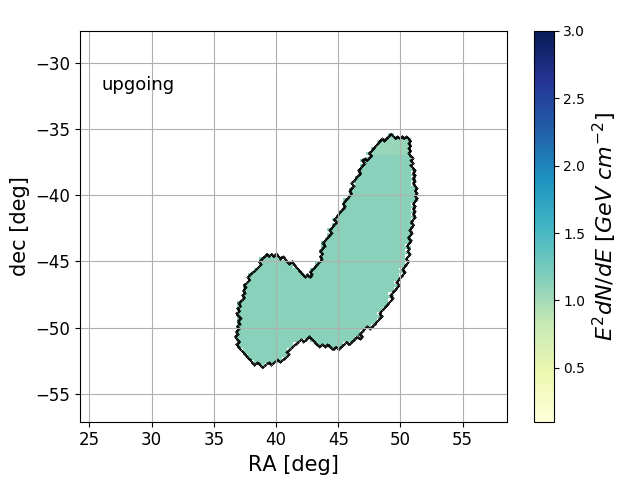}
    
    \hspace{-1.2cm}\includegraphics[scale=0.52]{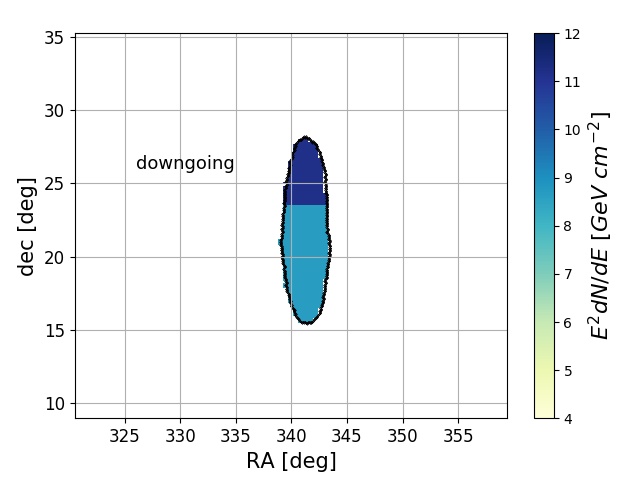}
    \includegraphics[scale=0.52]{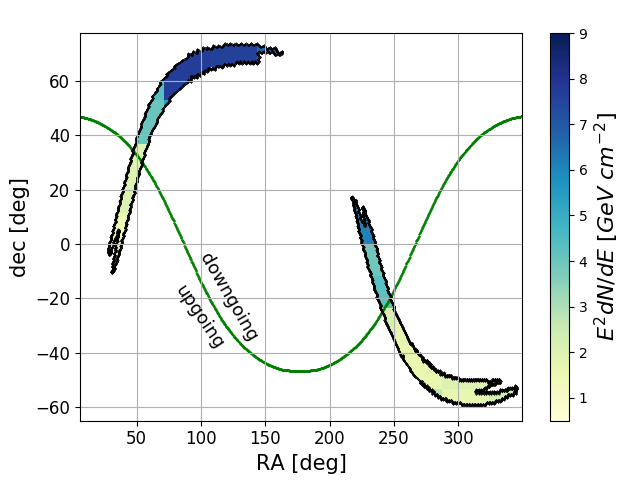}
    \vspace{-0.7cm}
    
    \caption{Upper limits on the neutrino spectral fluence (colored scale) as a function of the position in the sky in equatorial coordinates, computed assuming an E$^{-2}$ neutrino spectrum. The events are shown in chronological order: GW170608 (upper left), GW170729 (upper right), GW170809 (middle left), GW170814 (middle right), GW170818 (bottom left) and GW170823 (bottom right). The 90\% GW localization contour is superimposed. The green line indicates the ANTARES horizon, below the line corresponds to upgoing events and above the line to events above the horizon (downgoing).}
    \label{fig:limits}
\end{figure}

The upper limits obtained, assuming a neutrino spectrum with spectral index $\gamma$=2 (generic model typically expected for Fermi acceleration~\cite{fermiacc}), are shown in Fig.~\ref{fig:limits} for the six GW events. Table~\ref{table:limits} provides the average 90\% CL fluence UL ($\phi_{0}^{90\%}$) inside the error box.

\begin{table}[h!]
 \begin{center}
  \caption{Average fluence upper limit inside the 90\% CL contour, including tracks and showers.} 
   \vspace{0.1cm} 
  \label{table:limits}
 
 {
  \begin{tabular}{|c|c|c|}\hline
  {\bf GW event} &  \multicolumn{2}{c|}{ \bf $\phi_{0}^{90\%}$ (GeV cm$^{-2}$)} \\ \hline 
{} & {\hspace{0.2cm} Upgoing \hspace{0.2cm}} & {Downgoing} \\ \hline 
GW170608 &  1.6$\pm$0.5  & 2.2$\pm$0.9 \\ \hline 
GW170729 &  1.7$\pm$0.5  & 4.0$\pm$1.0 \\ \hline 
GW170809 &  1.1$\pm$0.3  & 1.2$\pm$0.5 \\ \hline 
GW170814 &  1.1$\pm$0.3  & - \\ \hline 
GW170818 &  -  & 9.0$\pm$4.0 \\ \hline 
GW170823 &  1.7$\pm$0.5  & 6.0$\pm$2.0 \\ \hline 
  \end{tabular}
  }
 \end{center}
\end{table}

The main systematic uncertainties on the estimated fluence upper limit come from two sources. The first is the uncertainty on the detector acceptance, which is related to the photon detection efficiency of the PMTs. It comprises an angular effect that leads to a 15\% variation of the flux of upgoing events~\cite{PS2012} and a 30\% variation for downgoing events~\cite{thesis}, and an overall effect due to the quantum efficiency of the PMTs and optical water properties, which results in a variation of about 20\% on the total acceptance~\cite{atmflux}. The second source relates to the capability of the MC simulation to reproduce data conditions on a run-by-run basis. It was evaluated together for upgoing and downgoing events and amount to $\sim$20\% \cite{GW170104}. For the shower event topology, an additional systematic effect of $\sim$7\% is introduced to account for the uncertainty on the shower position inside the GW error box. All the mentioned effects account for a total systematic uncertainty on the fluence upper limit of about 33\% for upgoing and 42\% for downgoing events.

\begin{table}[h!]
 \begin{center}
  \caption{Average upper limit inside the 90\% CL contour for the previous GW-neutrino follow-up analyses.}
   \vspace{0.1cm} 
  \label{t:summlim}
 {
  \begin{tabular}{|c|c|c|c|c|}\hline
  {\bf GW event} & {\bf sky region} & { \bf $\phi_{0}^{90\%}$ (GeV cm$^{-2}$)} \\ \hline
  GW150914 & IceCube downgoing & 1.2$\pm$0.3 \\ \hline
  GW151226 & IceCube upgoing & $\sim$0.06 \\ \hline
  GW151226 & IceCube downgoing & $\sim$0.9\\ \hline
  GW170104 & ANTARES upgoing & $\sim$2 \\ \hline
  GW170104 & ANTARES downgoing & $\sim$20 \\ \hline
  \end{tabular}
  }
 \end{center}
\end{table}\vspace{-0.4cm}

Table~\ref{t:summlim} provides a summary of the current neutrino limits from previous GW-neutrino follow-up analysis. The first three raws in the Table are two O1 events for which the analysis was performed independently by IceCube and ANTARES telescopes \cite{GW150914,GW151226}. For this two events, the value reported in the Table refers to the most sensitive detector. The third event (GW170104) is the only published event from O2 before this work. It is an ANTARES only analysis and considers events from the full sky \cite{GW170104}. For the second and third events, the values are extrapolated from the skymaps.

\subsection{Constraints on the total energy}

From the null detection and using the 90\% CL upper limit obtained in the previous section, a constraint on the total equivalent isotropic energy (E$_{\nu,{\rm iso}}$) emitted by the source in high-energy neutrinos, within the sensitive energy range of the search (TeV-PeV range), can be set. For this, the mean of the reconstructed luminosity distance distribution inside the error box provided by LIGO-Virgo~\cite{GWTC-1} is used for the redshift estimate. 

The total energy emitted in high-energy neutrinos is computed  according to Eq.~\ref{eq:limit} by integrating the neutrino spectrum over the energy range expected to contain 5-95\% of detected events (see Table~\ref{tab:eiso}) together with the measured luminosity distance and the associated redshift,

\begin{equation}
E_{\nu,{\rm iso} } = \frac{ 4 \pi D_L(z)^{2} }{ 1+z } \int^{E^{95\%}}_{E^{5\%}} E_{\nu}^{-2} \phi_0^{90\%} E_{\nu} dE_{\nu}.
\label{eq:limit}
\end{equation}

\begin{table}[!h]
 \begin{center}
  \caption{Average upper limit (E$_{\nu,{\rm iso}}^{\rm up/down}$) on the total neutrino energy emitted inside the 90\% confident area, computed within the 5-95\% energy range of the analysis, for upgoing and downgoing events and for each GW event. The measured redshift for each event is also provided.}
   \vspace{0.1cm} 
  \label{tab:eiso}
  \begin{tabular}{|c|c|c|c|c|c|}\hline   
{\bf GW event} & {\bf redshift (z) } &  { \bf E$_{5-95\%}^{\rm up}$} & { \bf E$_{5-95\%}^{\rm down}$} & {\bf E$_{\nu,{\rm iso}}^{\rm up}$ [erg]} & {\bf E$_{\nu,{\rm iso}}^{\rm down}$ [erg]} \\ \hline
GW170608 & 0.07$^{+0.02}_{-0.02}$ & 2.5~TeV - 4.0~PeV & 20~TeV - 25~PeV & 2.2$\times 10^{53}$ & 2.9$\times 10^{53}$ \\ \hline
GW170729 & 0.48$^{+0.19}_{-0.20}$ & 3.2~TeV - 4.0~PeV & 32~TeV - 25~PeV & 1.2$\times 10^{55}$ & 2.6$\times 10^{55}$ \\ \hline
GW170809 & 0.20$^{+0.05}_{-0.07}$ & 3.2~TeV - 4.0~PeV & 8~TeV - 20~PeV & 1.2$\times 10^{54}$ & 1.5$\times 10^{54}$ \\ \hline
GW170814 & 0.12$^{+0.03}_{-0.04}$ & 2.5~TeV - 5.0~PeV & - & 4.8$\times 10^{53}$ & - \\ \hline
GW170818 & 0.20$^{+0.07}_{-0.07}$ & - & 20~TeV - 32~PeV & - & 1.1$\times 10^{55}$ \\ \hline
GW170823 & 0.34$^{+0.13}_{-0.14}$ & 4.0~TeV - 4.0~PeV & 20~TeV - 25~PeV & 5.7$\times 10^{54}$ & 1.9$\times 10^{55}$ \\ \hline
  \end{tabular}
 \end{center}
\end{table}

\begin{figure}[!h]
\centering 
\vspace{-0.3cm}

\hspace{-0.3cm}\includegraphics[scale=0.54]{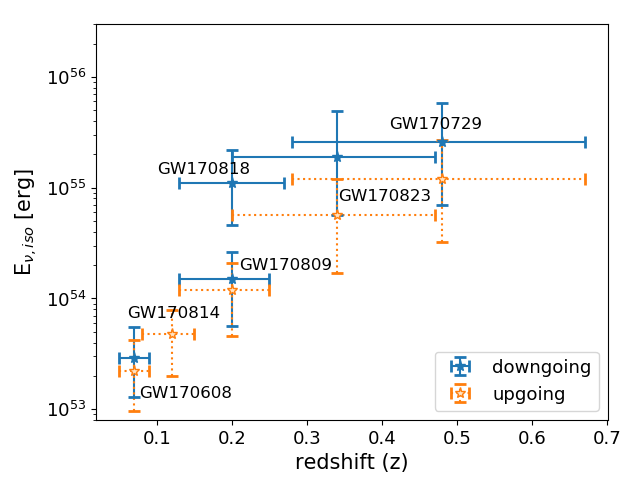}
\vspace{-0.4cm}

\caption{90\% CL upper limits on the total isotropic energy emitted in neutrinos within the 5-95\% energy range of the search for the six GW events analysed as a function of the estimated redshift. Results are given for the ANTARES downgoing (blue) and upgoing (orange) searches. The error bars in the X-axis correspond to the uncertainty on the distance estimate. The error bars in the Y-axis indicate the maximum and minimum limits obtained within this redshift range.}
\label{fig:eiso}
\end{figure}

The average 90\% CL upper limits inside the error box are summarised in Table~\ref{tab:eiso} together with the 5-95\% energy range for each GW event. In Fig.~\ref{fig:eiso}, these results are shown as a function of the redshift for six GW events, separately for the upgoing (orange) and downgoing (blue) regions of the error box, using the average limits inside the 90\% confidence regions and distances. As inferred from Eq.~\ref{eq:limit}, these limits scale with $D_{L}^{2}$ and proportionally to the fluence limits ($\phi_{0}^{90\%}$), which have been here obtained for a $E^{-2}$ neutrino spectrum. These limits depend on the background rates, which are mainly low and isotropic in local coordinates for the part of the error box below the horizon, while are highly dependent on the zenith angle for the part above the horizon.

\section{\label{s:conc} Conclusions}
A neutrino search using the ANTARES neutrino telescope data yields no neutrino observed in association with the six confirmed GW signals from the analysed Advanced-LIGO and Virgo during O2 run. From the null result, upper limits on the neutrino spectral fluence and on the isotropic energy radiated through neutrinos in the 5-95\% energy range of the search, to which the ANTARES detector is sensitive, are set for each GW analysed event.

This work presents an all-flavor neutrino analysis applied to GW events. The improvement on the sensitivity achieved by incorporating the shower topology compared to looking only at upgoing muon neutrinos has been estimated to range between 15\% to 30\% for the error box region yielding upgoing events, and up to a 200\% improvement for the region yielding downgoing events. This is due to the reduced background arising from atmospheric muons for the topology of contained showers. 

The Advanced-LIGO and Advanced-Virgo detectors are now taking data for a new scientific run, O3. Since the beginning of O3, 48 GW event candidates have been triggered up to the end of January 2020, and 31 of them are BBH merger candidates. The ANTARES Collaboration has released 46 GCN circulars with the follow-up real-time results. A stacking analysis of all BBH events from O3 is planned after the end of data taking by Advanced-LIGO/Virgo and the release of the GW catalogue. For the other astrophysical events, a similar refined analysis as the one presented here is aimed once they are confirmed. 

The ANTARES transient search method has been refined and shown to be robust for full-sky and all-flavour neutrino searches. Similar methods could be applied to localised flaring sources in the downgoing sky, where no analysis has been done aside from \cite{GW170817}. Indeed, this method has been adapted to the search for neutrino counterparts to Fast Radio Bursts \cite{FRBS} and it is being implemented for very high-energy Gamma-Ray Bursts observed by IACT's. Since these are localised sources, a better sensitivity due to the reduced background rate can be achieved. 

\section{Acknowledgements}
The authors acknowledge the financial support of the funding agencies:
Centre National de la Recherche Scientifique (CNRS), Commissariat \`a
l'\'ener\-gie atomique et aux \'energies alternatives (CEA),
Commission Europ\'eenne (FEDER fund and Marie Curie Program),
Institut Universitaire de France (IUF), LabEx UnivEarthS (ANR-10-LABX-0023 and ANR-18-IDEX-0001),
R\'egion \^Ile-de-France (DIM-ACAV), R\'egion
Alsace (contrat CPER), R\'egion Provence-Alpes-C\^ote d'Azur,
D\'e\-par\-tement du Var and Ville de La
Seyne-sur-Mer, France;
Bundesministerium f\"ur Bildung und Forschung
(BMBF), Germany; 
Istituto Nazionale di Fisica Nucleare (INFN), Italy;
Nederlandse organisatie voor Wetenschappelijk Onderzoek (NWO), the Netherlands;
Council of the President of the Russian Federation for young
scientists and leading scientific schools supporting grants, Russia;
Executive Unit for Financing Higher Education, Research, Development and Innovation (UEFISCDI), Romania;
Ministerio de Ciencia, Innovaci\'{o}n, Investigaci\'{o}n y Universidades (MCIU): Programa Estatal de Generaci\'{o}n de Conocimiento (refs. PGC2018-096663-B-C41, -A-C42, -B-C43, -B-C44) (MCIU/FEDER), Severo Ochoa Centre of Excellence and MultiDark Consolider (MCIU), Junta de Andaluc\'{i}a (ref. SOMM17/6104/UGR), 
Generalitat Valenciana: Grisol\'{i}a (ref. GRISOLIA/2018/119), Spain; 
Ministry of Higher Education, Scientific Research and Professional Training, Morocco.
We also acknowledge the technical support of Ifremer, AIM and Foselev Marine
for the sea operation and the CC-IN2P3 for the computing facilities.


\end{document}